\documentclass{article}

\pdfoutput=1 

\usepackage[square,sort,comma,numbers]{natbib}
\usepackage[preprint]{neurips_2023}

\usepackage[utf8]{inputenc} 
\usepackage[T1]{fontenc}    
\usepackage{hyperref}       
\usepackage{url}            
\usepackage{booktabs}       
\usepackage{amsfonts}       
\usepackage{nicefrac}       
\usepackage{microtype}      
\usepackage{xcolor}         
\usepackage[pdftex]{graphicx} 
\usepackage{tabularx}
\usepackage{textcomp}

\usepackage{amsmath,amssymb,amsfonts}
\usepackage{algorithmic}
\usepackage{caption,subcaption}

\title{Strengths and Weaknesses of 3D Pose Estimation and Inertial Motion Capture System for Movement Therapy}

\author{%
	Shawan Mohammed  \\
	Department of Electrical Engineering \\
	RWTH Aachen University\\
	Aachen, Germany \\
	\texttt{mohammed@ice.rwth-aachen.de} \\
	\And
	Hannah Siebers\\
	Department of Orthopaedics \\
	University Hospital RWTH Aachen\\
	Aachen, Germany \\
	\texttt{hsiebers@ukaachen.de} \\
	\AND
	Ted Preuß \\
	Department of Electrical Engineering\\
	RWTH Aachen University\\
	Aachen, Germany \\
	\texttt{sebastian.preuss@rwth-aachen.de} \\
}

\begin{document}

	\maketitle

	\begin{abstract}
3D pose estimation offers the opportunity for fast, non-invasive, and accurate motion analysis. This is of special interest also for clinical use. 
Currently, motion capture systems are used, as they offer robust and precise data acquisition, which is essential in the case of clinical applications. 
In this study, we investigate the accuracy of the state-of-the-art 3D position estimation approach MeTrabs, compared to the established inertial sensor system MTw Awinda for specific motion exercises. 
The study uses and provides an evaluation dataset of parallel recordings from ten subjects during various movement therapy exercises. 
The information from the Awinda system and the frames for monocular pose estimation are synchronized. 
For the comparison, clinically relevant parameters for joint angles of ankle, knee, back, and elbow flexion-extension were estimated and evaluated using mean, median, and maximum deviation between the calculated joint angles for the different exercises, camera positions, and clothing items. 
The results of the analysis indicate that the mean and median deviations can be kept below 5° for some of the studied angles. 
These joints could be considered for medical applications even considering the maximum deviations of 15°. 
However, caution should be applied to certain particularly problematic joints. 
In particular, elbow flexions, which showed high maximum deviations of up to 50° in our analysis. 
Furthermore, the type of exercise plays a crucial role in the reliable and safe application of the 3D position estimation method. 
For example, all joint angles showed a significant deterioration in performance during exercises near the ground.
	\end{abstract}

\section{Introduction}
Motion analysis has long been central to clinical and sports assessment, rehabilitation, and research \cite{GAIT1}, \cite{GAIT2}. The ability to quantitatively capture and analyze movement patterns is essential to diagnose disorders, monitor the effectiveness of therapeutic interventions, and develop biomechanical models to predict and improve performance \cite{INI2005}.
The development of noninvasive, high-precision motion analysis systems has made great strides in recent years \cite{wii2010}, \cite{GAIT3}, \cite{GAIT4}. Traditionally, inertial sensor systems, such as the MTw Awinda, have played an essential role in clinical and sports motion analysis \cite{awinda}, \cite{awinda2}. These systems provide robust and accurate acquisition of motion data, which is essential in clinical contexts. However, they also come with several limitations, including high cost, the complexity of calibration, and the significant time required to attach the soft tissue artifacts to the subjects, limiting their use in certain situations.

Deep Learning approaches for motion analysis, such as 3D position estimation, have gained increasing interest in recent years \cite{metrabs}, \cite{3dpose}. These technologies offer the ability to acquire motion data quickly, cheaply, and also non-invasively by applying deep neural network-based algorithms to images or video sequences. One promising method in this area is MeTrabs \cite{metrabs}, which aims to estimate 3D poses from monocular images. It addresses several limitations of conventional 2.5D methods that require separate post-processing to resolve scale ambiguities. In particular, MeTrabs can localize body joints outside of the image boundaries, allowing for complete estimation even in cropped images. These features make MeTrabs particularly suitable for use in our evaluation.
Furthermore, MeTrabs is based on the concept of volumetric heatmaps whose dimensions are all defined in 3D metric space rather than being aligned with image space. This reinterpretation of heatmap dimensions allows complete, metric scale poses to be estimated directly at test time without knowing distance or using anthropometric heuristics such as bone lengths. However, the performance of a 3D pose estimation is measured by several metrics, including Mean Per Joint Position Error (MPJPE). This metric evaluates how accurately the model predicts joint positions in 3D space compared to ground truth joint positions that was collected with gold standards like marker-based 3D motion capture systems \cite{markerbased}. The performance of a 3D pose estimation is commonly evaluated using various metrics, including Mean Per Joint Position Error (MPJPE). This metric evaluates the precise agreement of the joint positions in 3D space predicted by the model with the actual joint positions determined by marker-based 3D motion capture systems. In the context of these benchmark datasets, Metrabs achieved quite remarkable results. Specifically, on the Human3.6M dataset \cite{h36m_pami}, Metrabs achieved an average MPJPE of 49.3 mm with a standard deviation of ±0.7 mm. On the MPI-INF-3DHP dataset \cite{3dhp2017}, Metrabs recorded an average MPJPE of 74.9 mm with a standard deviation of ±1.4 mm. These results highlight the strong performance of Metrabs compared to other 3D pose estimation methods. Although this method has made impressive progress in recent years, its performance in the context of motion therapy remains unclear.
\paragraph{Objective} This study aims to present a detailed comparative analysis between MeTrabs and Awinda for different motion exercises. The evaluation of MeTrabs is done with respect to the Awinda system, as the latter is an established benchmark due to its wide application and recognition in research and practice. The Awinda inertial sensor system has a proven track record of high accuracy, robustness, and reliable data acquisition, making it a preferred tool for motion analysis. Despite potential errors in data acquisition and processing that can occur in the Awinda system, its comparatively low error rate and repeatable results have made it a widely accepted "tool" in the industry. In contrast to human beings observing and analyzing motion patterns affected by their experience in doing, stress level, emotions, and visual capability, the Awinda system allows standardized objective data acquisition for the description of motion patterns.
Using Awinda as a basis for comparison provides a valuable benchmark for newer technologies such as MeTrabs. MeTrabs, on the other hand, represents a new generation of motion capture systems that offers significant advantages, such as ease and speed of use and a low price through AI and Deep Learning. Despite these advantages, however, questions remain regarding its accuracy and robustness for clinical use. 

\paragraph{Contribution} This study highlights the limitations and challenges of 3D pose estimation in movement therapy and clinical applications. In particular, we focus on the role of training data, which is essential for the performance of these data-driven approaches. It is widely recognized that many weaknesses of 3D pose estimation result from limited or inadequate training data. Accordingly, we highlight specific accuracy shortcomings of 3D pose estimation to illustrate the type of training data required to enable robust clinical analysis. Compared to established systems such as Awinda, we aim to better understand and quantify the strengths and weaknesses of MeTrabs in relation to exercise therapy. Our ultimate goal is to evaluate whether MeTrabs can act as a viable alternative to Awinda by testing its accuracy and robustness in a real-world exercise therapy scenario.
The findings from this study should help optimize the use of advanced 3D pose estimation approaches in motion therapy and provide valuable insights for the future development and application of these technologies in clinical practice. The frameworks used for this study are all open source and free to use, including a framework for estimating posture using cameras \cite{mTower} and a user interface app \cite{hygym}.
\section{Related Work}
The analysis and evaluation of human motion have been a topic of continued interest in the medical, sports, and research communities \cite{GAIT1}, \cite{INI2005}. Over the years, several approaches have been developed and refined to capture and understand human movement more accurately \cite{wii2010}.
Traditional approaches rely primarily on marker-based systems such as Qualisys Miqus, known for their robust and accurate motion data acquisition \cite{qualisys}. Inertial sensor systems such as MTw Awinda, which are less accurate but more flexible in their application, are also well-known alternatives \cite{awinda}, \cite{awinda2}.
However, they come with notable limitations, including high cost, complex calibration, and significant time requirements for setup \cite{GAIT3}, \cite{GAIT2}. To overcome these limitations, noninvasive and computationally efficient methods are being developed and explored.
One such direction is depth sensors, as in the Kinect system \cite{kinect}. Although these systems present an affordable and reasonably accurate method for motion capture, the accuracy is generally lower than that of marker-based or inertial sensor systems. This trade-off between cost and accuracy is essential in practical applications, particularly in interactive environments such as exergaming for balance training. Vonstad et al. \cite{evalPoseEs2} explored this balance by comparing a deep learning-based pose estimation system with marker-based and Kinect systems, providing important insights for researchers and practitioners alike.
Recently, the field has seen rapid advancements with the emergence of deep learning approaches for motion analysis, such as 3D position estimation \cite{overViewPoseEs}. One promising method is MeTrabs, which applies deep neural network-based algorithms to images or video sequences to estimate 3D poses from monocular images \cite{metrabs}. These methods have demonstrated the potential to quickly, cheaply, and non-invasively capture accurate motion data.
However, the effective use of these systems relies heavily on accurate data acquisition. For example, OpenPose \cite{OpenPose}, a commonly used system for markerless motion capture, has been evaluated for its accuracy in various settings. Nakano et al. \cite{evalPoseEs1} focused on assessing the accuracy of OpenPose with multiple video cameras, underlining the importance of the environment and data acquisition setup in motion analysis studies.

\section{Methodology}
This section explains the technologies used and our analysis methodology between the Awinda inertial sensor system and the MeTrabs system for 3D position estimation. The goal is to determine the performance and accuracy of both systems, focusing on their use in clinical and therapeutic contexts. The methodological approach considers the challenges of 3D position estimation and the role of training data.
\subsection{Awinda Inertial Sensor System}
The Awinda inertial sensor system, manufactured by Movella, is a high-precision wireless motion capture system. It is based on a combination of inertial sensors and a magnetic field sensor system. The system is known for its high accuracy, robustness, and flexibility and is widely used in clinical and sports motion analysis.
The Awinda system consists of several main components. These include several small, lightweight sensors that are placed on the subject's body to collect motion data. The sensors were attached in strict compliance with the manufacturer's instructions and individual body characteristics of the subjects. Elastic Velcro straps were used to ensure a comfortable fit and optimal freedom of movement.
Of crucial importance for the quality of the acquired data is the calibration process. Here, we ensured that this process was performed carefully and without interference. The subjects performed a series of standard movements to adjust the system to their specific movement patterns.
To ensure calibration accuracy, we did not rely solely on the calibration algorithms provided by Awinda. In addition, we verified that the skeletal structure captured by the sensors and the actual body posture of the subjects matched. Especially at the arms, elbows, and ankles, this was important to detect possible errors in sensor placement and calibration.

\subsection{3D Position Estimation with MeTrabs}
\begin{figure}[!h]
	\begin{center}
		\includegraphics[width=0.8\linewidth]{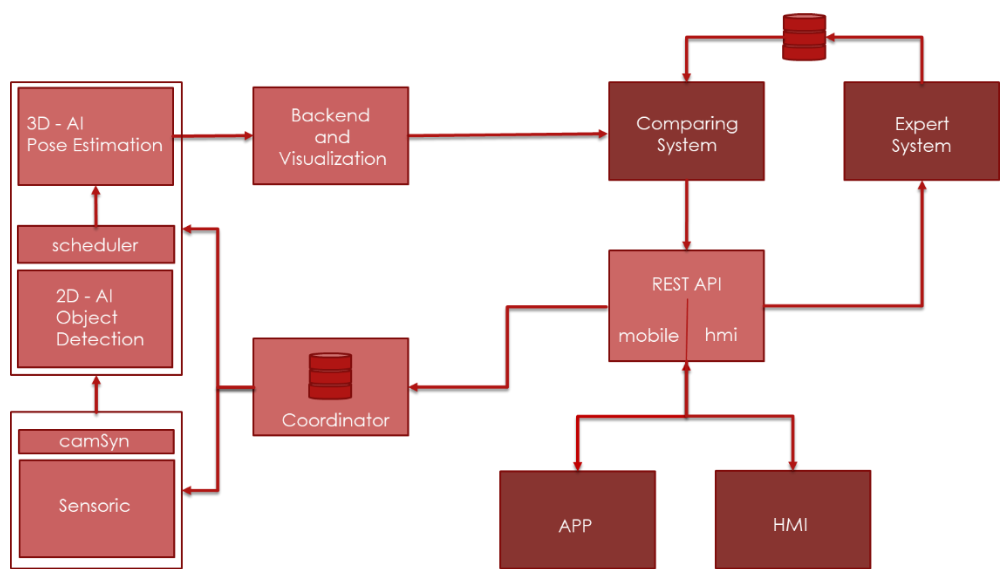}
		\caption{Architectural design of the ROS frame.}
		\label{fig:ros_plattform}
	\end{center}
\end{figure}

\begin{figure}[!h]
	\begin{center}
		\includegraphics[width=0.8\linewidth]{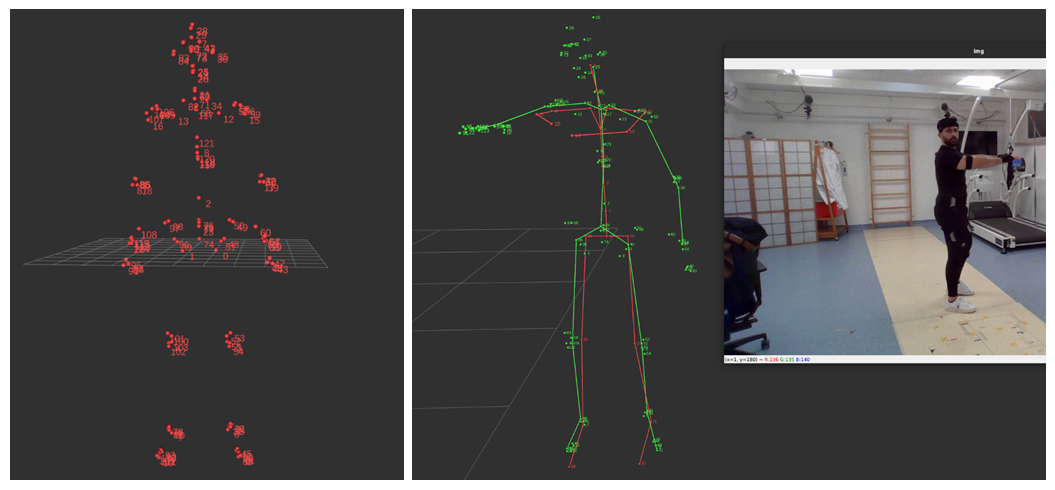}
		\caption{Comparison of MeTrabs and Awinda skeleton in ROS framework.}
		\label{fig:comparison}
	\end{center}
\end{figure}

In the context of our study, the MeTrabs method was used to determine body postures. Rather than using traditional metrics such as Mean Per Joint Position Error (MPJPE), we calculated and analyzed angles at four specific joints: trunk, elbow, knee, and ankle. These angles were calculated by analyzing MeTrabs skeletal data and comparing them to the Awinda inertial sensor system values.
In parallel with the data acquisition with Awinda, we used a video camera to capture the subjects' movements. The official Awinda software ensured the synchronization of video and Awinda data, allowing the precise temporal matching of video frames and motion data.
The individual frames from the video recordings were subsequently used to pedicure 17 joints with MeTrabs, which formed a skeleton. By using this technique, we could carefully verify the accuracy of MeTrabs' 3D position estimation and directly compare it to the results from the Awinda system.
The individual video frames were analyzed using MeTrabs, locating 17 joints in each frame and modeling a resulting skeleton. This method allowed us to examine in detail the precision of MeTrabs' 3D position estimation and to contrast it directly with the results of the Awinda system.
In addition, the MeTrabs model was integrated into a framework \cite{mTower} using the Robot Operating System (ROS) as middleware \cite{ROS22}, see Figure \ref{fig:ros_plattform}. This configuration allowed the normalization of skeletal data from both MeTrabs and Awinda to produce a coherent scale for systematic comparison.
Furthermore, the framework allowed the skeletons from both systems to be overlaid to ensure congruence. This congruence is significant, especially given that the underlying structure of the MeTrabs and Awinda skeletons varies. Such differences are common in 3D position determination, as skeletal structures differ even in state-of-the-art datasets. Since we cannot modify the structure of MeTrabs and Awinda, normalization and subsequent overlay of these structures are essential.
The developed ROS framework also allows real-time visualization of the skeletons of MeTrabs and Awinda using Rviz. This feature facilitated an immediate comparison of the skeletal visualizations of both systems and contributed to a deeper understanding of their differences, see Figure \ref{fig:comparison}.

The real-time execution capability of the ROS framework ensured continuous monitoring and adjustment of the analysis process. Complementing this, an API interface was implemented to allow access and control of the framework from various endpoints, including smartphone applications \cite{hygym}, facilitating the retrieval of relevant data.

\paragraph{Mapping and Normalization}
A significant issue leading to initial deviations and the introduction of an offset error is the substantial difference between the skeletons of MeTrabs and Awinda. The initial implementation of MeTrabs allows using 128 virtual joint points beyond the 17 joints originally predicted by MeTrabs neural network, see Figure 2. From this selection of extra joint points, we can assemble a skeletal structure that provides a better degree of coverage with Awinda and thus produce more meaningful results.
In addition, we implemented a mapping process to overlay the two skeletons to minimize the error between the joints. In this process, we align the two skeletons to minimize the Euclidean distance between the joint points at the shoulders and hips.
The final step in our methodology is to normalize the skeletons to a uniform size and center them around the coordinate center. This is done to compensate for differences due to the individual body sizes of the subjects, thus providing a more comparable basis for data analysis.
Throughout the post-processing of the Metrabs skeleton, we were always careful to match the two skeletons as closely as possible to create a more solid basis for comparison. In doing so, it was important not to falsify the original predictions. Using this approach, we can minimize the bias between the skeleton models while maintaining the integrity of the original data.

\subsection{Comparative Analysis of Awinda and MeTrabs:}
The focus of our study is to directly compare the data generated by the two systems in terms of joint angle acquisition during different motion exercises.
The analysis method is based on simultaneous data acquisition from both systems during various motion exercises. These exercises were carefully selected to cover various movement types commonly encountered in clinical practice and movement therapy.
These movement data from multiple subjects were carefully generated, synchronized, and stroked into a single data set. Subjects performed various movement exercises, including sit-ups, push-ups, and squats. These exercises were chosen for their variance in body orientation - they included upright, prone, and bent positions. Prone exercises, in particular, present challenges for both systems. Especially 3D Pose Estimation approaches, like MeTrabs, brings it to its limits, as also shown in the original work of MeTrabs regarding the evaluation on Human3.6M \cite{metrabs}.
Each subject performed each exercise from a given perspective precisely ten times to ensure a sufficient amount of data for analysis and to allow statistically significant results. For Awinda, the inertial sensors were placed at key locations on the subjects' bodies, while MeTrabs captured the subjects' movements via a camera pointed at them. Subjects performed the exercises from two different perspectives to the camera: once at 45° and once frontal (0°) to the camera. While the Awinda system operates independently of these configurations, the positioning of the subject relative to the camera has a significant impact on data acquisition in the MeTrabs system, as more or fewer joints may be occluded depending on the position and exercise, making them more difficult to capture by the MeTrabs module.
The analysis focuses on four specific joints and always examines flexion only: the left and right knee, left and right ankle, back (measured in two different ways and compared to the Awinda $vertical\_T8$ and $vertical\_Pelvis$ angles), and left and right elbow.
The Awinda system calculates the raw angles based on the different IMU sensors and allows access to this data via the Awinda software. In addition, Awinda provides other evaluations and information, such as a skeleton of the subject.
In contrast, MeTrabs does not provide angles of the different skeletal elements but only a skeleton consisting of 17 joint points. This data is used to calculate the joint angles, which are then compared with the values output directly by Awinda.

\begin{figure}[!h]
	\begin{center}
		\includegraphics[width=0.9\linewidth]{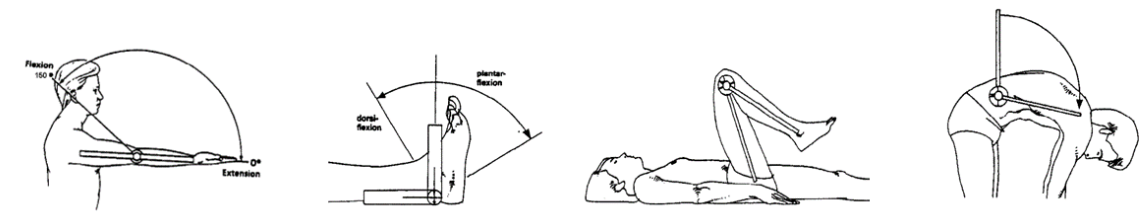}
		\caption{The flexion is calculated in the plane for knee, ankle, back segment and elbow. \cite{physiothera}}
		\label{fig:joints}
	\end{center}
\end{figure}

\paragraph{Calculation basis of the four joints}
The following explanation highlights the approach to calculating the four specific joints we consider in this study:
Our calculation method is based on sound physiotherapeutic knowledge and considers the human body's biomechanics. It is a fundamental principle of human biomechanics that specific movements are allowed or limited based on the physical structure and function of the body. In order to quantify this variety of movements in different directions, a systematic approach is needed that considers the specific characteristics of each joint and the type of angle to be measured.
The four joints we analyzed - the knees, the ankles, the back segment, and the elbows - are considered hinge joints in our calculation and physiotherapy in general. This means that we only assume a movement within a plane defined by two axes for flexion measurement, see Figure \ref{fig:joints}.
However, before we go deeper into the angle calculation method, we must clarify the definition of joint angles and the defined axes in the Awinda technology. The angles provided by the Awinda system are defined in a specific coordinate system that allows precise measurement of movements in three-dimensional space.
Metrabs, on the other hand, works based on camera coordinates instead of global world coordinates, which means that the x,y, and z are in relation to the camera.
To finally compare Awinda and Metrabs, the Metrabs 3D coordinates have to be adjusted to the Awinda world coordinates. 
For this, we used the Euler rotation, which is also used internally by Awinda to transform uniformly between the coordinates of the individual sensors and a uniform world coordinate.
Once the axes are appropriately aligned, we continue calculating the flexion angles based on considering the joints as hinge joints. Here, one of the three axes is neutralized, creating a two-dimensional plane. In this context, we define flexion as the motion within this plane that describes the position of the joint with respect to the remaining two axes. This method allows us to quantify and compare joint motion accurately.

\section{Results}
In our results section, we look at the eight different flexion angles derived from the four joints considered, compared to the angles provided by the Awinda system. We are looking at the differences between the predicted angles. In addition to plotting these differences over the time axis in seconds, we provide other statistical characteristics, including maximum, median, and average deviation.
To evaluate the accuracy of the algorithms we use - specifically mapping, normalization, and converting 3D joint positions to specific angles. Therefore, we first apply them to data where we expect high accuracy. This is done by comparing the raw angles predicted by Awinda to the angles we computed based on the Awinda skeleton.
Through this process, we can identify the algorithmic error as well as possible error offsets that occur due to internal calculations, assumptions, and corrections used by the Awinda system that we are not aware of. This approach lets us get a clear picture of the accuracy of the algorithms we use for analysis while identifying potential areas where the Awinda system has weaknesses or anomalies in its data.

Table \ref{tab:algoSquats} describes the maximum, median, and average deviation for the algorithmic deviation between the raw Awinda angles and the angles calculated based on the Awinda skeleton. The exercises evaluated in Table Table \ref{tab:algoSquats} consisted of squats observed across all subjects and all repetitions performed. From the average and median, it can be seen that the overall deviation is relatively small, and even the maximum deviations are only about 1° for most joints. However, there are isolated measurement points that show higher deviations. For example, we can observe this at the elbows with maximum deviations of 4 to 6°.
We explain these larger deviations by internal calculations, which are carried out within the Awinda system and are unknown to us. These aspects are not discussed in detail in scientific papers or the official documentation. The official documentation only points out certain conditions in sensor mounting that may lead to some offset compared to optical systems. Also, it is known that Awinda can predict relatively accurate angles with the IMUs, but this is not transferable to the skeleton predicted by Awinda. This is based on estimates of each subject's segments and input body parameters. Nevertheless, we cannot rule out that it is not due to our algorithms used.

\begin{table}
	\centering
	\scriptsize
	\caption{Awinda angle vs angle calculated using Awinda skeleton }
	\label{tab:algoSquats}
	\begin{tabular}{lll}
		\\
		\toprule
		\toprule
		\toprule
		Joint angle flexion \\
		\midrule
		Knee (right)	\\
		Knee (left)	\\
		Ankle (right)	\\
		Ankle (left)	\\
		Back (pelvis)	\\
		Back (T8)	\\
		Elbow (right)	\\
		Elbow (left)	\\
		\bottomrule
	\end{tabular}
	\begin{tabular}{lll}
		\toprule
		\multicolumn{3}{c}{Squats} \\
		\cmidrule(r){1-3}
		median [°] & average [°] &  maximum [°] \\
		\midrule
		0.11 & 0.11 & 0.30\\
		0.32 & 0.36 & 1.13\\
		0.61 & 0.62 & 1.23\\
		0.47 & 0.49 & 0.83\\
		0.01 & 0.24 & 0.07\\
		0.23 & 0.62 & 0.54\\
		0.36 & 1.16 & 6.17\\
		0.09 & 0.48 & 4.29\\
		\bottomrule
\end{tabular}
\end{table}

In Table \ref{tab:algoGroundNear}, similar to Table \ref{tab:algoSquats}, we present the algorithmic errors. These show higher deviations overall, consistent with our observations that the Awinda skeleton has difficulty in near-ground exercises. Nevertheless, the average and median errors remain moderate and do not exceed 7°.
These higher errors are due to occasional deviation peaks during exercise. In particular, the error for the back (T8) during sit-ups is highly noticeable, with a deviation of 41°. 

\begin{table}
	\centering
	\scriptsize
	\caption{Awinda angle vs angle calculated using Awinda skeleton }
	\label{tab:algoGroundNear}
	\begin{tabular}{lll}
		\\
		\toprule
		\toprule
		\toprule
		Joint angle flexion \\
		\midrule
		Knee (right)	\\
		Knee (left)	\\
		Ankle (right)	\\
		Ankle (left)	\\
		Back (pelvis)	\\
		Back (T8)	\\
		Elbow (right)	\\
		Elbow (left)	\\
		\bottomrule
	\end{tabular}
	\begin{tabular}{lll}
		\toprule
		\multicolumn{3}{c}{Situps} \\
		\cmidrule(r){1-3}
		median [°] & average [°] &  maximum [°] \\
		\midrule
		0.002 & 0.003 & 0.02\\
		0.001 & 0.003 & 0.16\\
		4.67 & 4.57 & 6.86\\
		3.16 & 3.30 & 4.83\\
		0.11 & 0.13 & 0.43\\
		0.12 & 0.16 & 0.61\\
		4.83 & 4.67 & 9.87\\
		1.70 & 1.67 & 5.13\\
		\bottomrule
	\end{tabular}
	\begin{tabular}{lll}
		\toprule
		\multicolumn{3}{c}{Pushups} \\
		\cmidrule(r){1-3}
		median [°] & average [°] &  maximum [°] \\
		\midrule
		0.01 & 0.02 & 0.17\\
		0.01 & 0.01 & 0.04\\
		0.71 & 0.70 & 1.21\\
		5.08 & 5.62 & 18.23\\
		2.38 & 6.84 & 12.09\\
		0.90 & 0.71 & 41.00\\
		2.83 & 2.83 & 5.24\\
		2.24 & 2.15 & 3.23\\
		\bottomrule
	\end{tabular}
\end{table}

Upon closer examination, this can be attributed to a systematic error in the Awinda system. This error is shown in Table \ref{tab:ErrorAwinda}: There is a sudden rotation of the entire z-axis by 180 degrees within only 15 milliseconds. According to our observations, this phenomenon occurs only at values below 90°.
We cannot fully explain this error, but we suspect it has something to do with the gravity vector. This is estimated and used together with the orientation of the basin to determine the vertical reference. Once this vertical reference is determined, the orientation of the basin is calculated with respect to this vertical frame to obtain the angular output.

\begin{table}
	\centering
	\scriptsize
	\caption{Anomaly in the Awinda angle data}
	\label{tab:ErrorAwinda}
	\begin{tabular}{lll}
		\toprule
		Time sequence No.\\
		\midrule
		1	\\
		2	\\
		3	\\
		4	\\
		5	\\
		6	\\
		7	\\
		8	\\
		\bottomrule
	\end{tabular}
	\begin{tabular}{lll}
		\toprule
		x [°] & y [°] &  z [°] \\
		\midrule
		67.41 & 0.00 & -80.14\\
		78.76 & 0.00 & -80.76\\
		88.37 & -180.00 & 99.41\\
		76.19 & 180.00 & 100.01\\
		67.74 & -180.00 & 100.70\\
		61.37 & 180.00 & 101.43\\
		56.43 & -180.00 & 102.16\\
		52.25 & -180.00 & 102.90\\
		\bottomrule
	\end{tabular}
\end{table}

Having examined the algorithmic error and weaknesses of the Awinda skeleton, it remains to be noted that all deviations are moderate and systematic, which qualifies it for scientific evaluation despite certain limitations. In the next step of our analysis, we no longer compare the Awinda skeleton but turn our focus to the MeTrabs skeleton, which we contrast with the angles provided by the Awinda system.

Table \ref{tab:MetrabsSquats} presents the maximum, median, and average deviation between the raw Awinda angles and those calculated based on the Metrabs skeleton. This analysis includes all squat exercises performed by all subjects and across all repetitions. The back, ankle, and knee flexions show average and median deviations between 1° and 6.69°, while the maximum deviations vary between 7.70° and 20.32°. It is noticeable that the side of the subjects facing the camera (left) always shows minor deviations on average. For the elbows, we notice more significant deviations from the Awinda angles; this observation could also be confirmed empirically.

\begin{table}
	\centering
	\scriptsize
	\caption{Awinda angle vs angle calculated using MeTrabs skeleton }
	\label{tab:MetrabsSquats}
	\begin{tabular}{lll}
		\\
		\toprule
		\toprule
		\toprule
		Joint angle flexion \\
		\midrule
		Knee (right)	\\
		Knee (left)	\\
		Ankle (right)	\\
		Ankle (left)	\\
		Back (pelvis)	\\
		Back (T8)	\\
		Elbow (right)	\\
		Elbow (left)	\\
		\bottomrule
	\end{tabular}
	\begin{tabular}{lll}
		\toprule
		\multicolumn{3}{c}{Squats} \\
		\cmidrule(r){1-3}
		median [°] & average [°] &  maximum [°] \\
		\midrule
		5.83 & 5.86 & 13.85\\
		4.38 & 6.11 & 20.32\\
		3.03 & 3.43 & 9.02\\
		1.27 & 1.89 & 7.70\\
		3.61 & 5.21 & 16.35\\
		6.69 & 6.51 & 13.33\\
		31.72 & 32.47 & 44.33\\
		21.58 & 23.38 & 49.01\\
		\bottomrule
	\end{tabular}
\end{table}

\begin{figure*}
	\makebox[\textwidth][c]{%
		\begin{subfigure}[b]{.21\paperwidth}
			\centering
			\includegraphics[width=.95\textwidth]{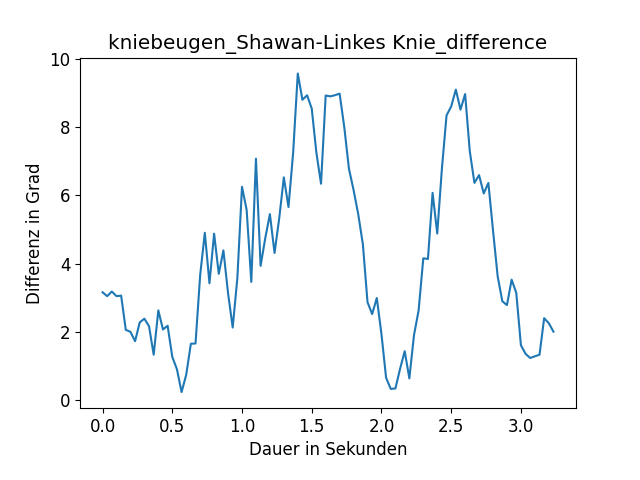}
			\caption{Knee}
			\label{fig:Knee}
		\end{subfigure}%
		\begin{subfigure}[b]{.21\paperwidth}
			\centering
			\includegraphics[width=.95\textwidth]{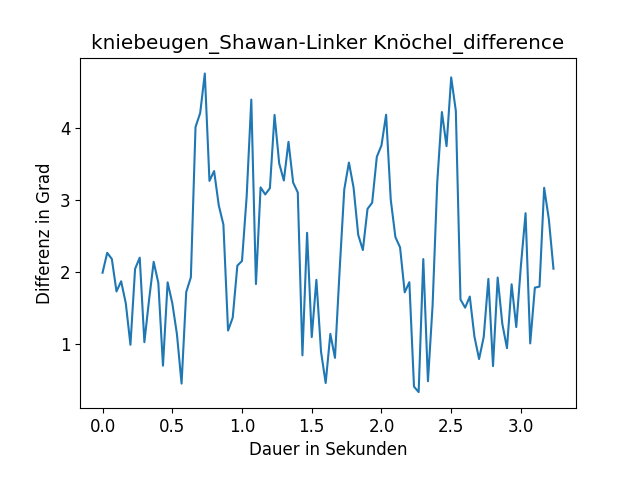}
			\caption{Ankle}
			\label{fig:ankle}
		\end{subfigure}%
		\begin{subfigure}[b]{.21\paperwidth}
			\centering
			\includegraphics[width=.95\textwidth]{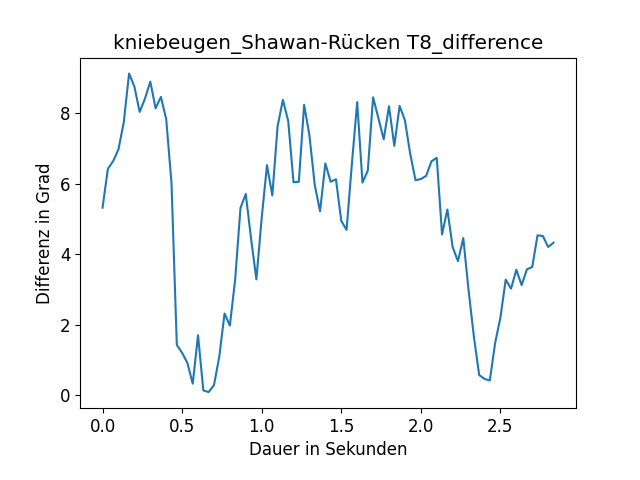}
			\caption{Back $T8$}
			\label{fig:back}
		\end{subfigure}%
	}
	\caption{Evaluation and comparison of different approaches based on mean cumulative reward per episode.}
\end{figure*}

Figures \ref{fig:Knee}, \ref{fig:ankle}, and \ref{fig:back} show the angular differences of the back, knee, and ankle angles over time. Here, an example repetition is shown and does not consider all repetitions over all available test persons as the evaluations in the tables do. This also explains why the angle differences in the figures show fewer deviations.

In Table \ref{tab:GroundNear}, we present the evaluation of the Metrabs skeleton for exercises near the ground (push-ups and sit-ups). As expected, all exercises and all studied joints show significantly worse results. These values are also significantly elevated compared to the algorithmic errors in Table \ref{tab:algoGroundNear}, highlighting the challenges of using the Metrabs skeleton for near-ground exercises.

\begin{table}
	\centering
	\scriptsize
	\caption{Awinda angle vs angle calculated using Awinda skeleton}
	\label{tab:GroundNear}
	\begin{tabular}{lll}
		\\
		\toprule
		\toprule
		\toprule
		Joint angle flexion \\
		\midrule
		Knee (right)	\\
		Knee (left)	\\
		Ankle (right)	\\
		Ankle (left)	\\
		Back (pelvis)	\\
		Back (T8)	\\
		Elbow (right)	\\
		Elbow (left)	\\
		\bottomrule
	\end{tabular}
	\begin{tabular}{lll}
		\toprule
		\multicolumn{3}{c}{Situps} \\
		\cmidrule(r){1-3}
		median [°] & average [°] &  maximum [°] \\
		\midrule
		9.61 & 9.32 & 14.04\\
		7.82 & 8.25 & 21.49\\
		4.67 & 5.54 & 15.45\\
		7.43 & 8.65 & 23.34\\
		21.75 & 22.42 & 29.44\\
		27.37 & 24.91 & 33.21\\
		43.94 & 45.73 & 73.02\\
		11.55 & 14.39 & 32.42\\
		\bottomrule
	\end{tabular}
	\begin{tabular}{lll}
		\toprule
		\multicolumn{3}{c}{Pushups} \\
		\cmidrule(r){1-3}
		median [°] & average [°] &  maximum [°] \\
		\midrule
		2.21 & 2.65 & 8.35\\
		5.12 & 5.19 & 13.94\\
		11.08 & 10.29 & 18.16\\
		16.58 & 14.34 & 29.17\\
		21.03 & 22.13 & 36.01\\
		12.33 & 16.47 & 53.54\\
		13.14 & 12.21 & 29.01\\
		4.97 & 6.30 & 38.75\\
		\bottomrule
	\end{tabular}
\end{table}

\section{Conclusion}
This study provided important insights into modern 3D position estimation performance in exercise therapy and demonstrated the potential for future medical applications. Weaknesses of 3D position estimation for medical applications include the high variances in near-ground motion. Also, certain joints are particularly susceptible to inaccuracies and should not be used as the basis of analysis for medical applications. The elbow joint, in particular, is the problem here, which may be a reason for this due to its high variability in data sets.
Furthermore, we could prove that the camera perspective has a decisive influence on the accuracy of the estimates. This aspect should be taken into account when designing applications.
Another important finding concerns the significant difference between standard exercises, such as standing exercises, and more specific 'corner case' exercises, such as exercises near the ground. This finding highlights the significant need for diversified datasets covering various movements and postures. Thus, 'in-the-wild' datasets \cite{3dhp2017} showing people in different environments, poses, and movement patterns should gain considerable importance and be further focused in 3D pose estimation research. In addition, there is an opportunity to selectively augment datasets to address previously identified vulnerabilities. Data-driven approaches such as pose estimation can be significantly increased in robustness and accuracy through subsequent training.
In conclusion, 3D pose estimation offers promising opportunities for medical motion therapy, but significant improvements are still needed to achieve its full effectiveness in this field.

\paragraph{Future Work}
For future work, the evaluation must be extended to include optical and marker-based approaches, as these methods are considered gold standards in motion capture. This becomes particularly relevant due to the discrepancies between the angles Awinda predicted and the skeleton generated by Awinda.
In addition, the number of exercises and joints to be examined should be expanded to cover a broader range of motion scenarios. This would allow for a more comprehensive assessment of the system's performance and could help to gain further insight into potential weaknesses and areas for improvement.
In addition to expanding the scope of exercises and joints, the variance of recording locations should also be considered. Varying environmental conditions can significantly impact the performance of Deep Learning-based computer vision applications, such as object detection using a camera \cite{noise}, \cite{Noise2}. Therefore, it is significant to consider these factors in future evaluations.

\small
\bibliography{./bib/references}	
	
\end{document}